# scientific reports

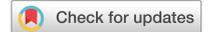
Check for updates

**OPEN** 
# A quantum graph approach to metamaterial design

Tristan Lawrie[1], Gregor Tanner[1✉] & Dimitrios Chronopoulos[2]

Since the turn of the century, metamaterials have gained a large amount of attention due to their potential for possessing highly nontrivial and exotic properties—such as cloaking or perfect lensing. There has been a great push to create reliable mathematical models that accurately describe the required material composition. Here, we consider a quantum graph approach to metamaterial design. An infinite square periodic quantum graph, constructed from vertices and edges, acts as a paradigm for a 2D metamaterial. Wave transport occurs along the edges with vertices acting as scatterers modelling sub-wavelength resonant elements. These resonant elements are constructed with the help of finite quantum graphs attached to each vertex of the lattice with customisable properties controlled by a unitary scattering matrix. The metamaterial properties are understood and engineered by manipulating the band diagram of the periodic structure. The engineered properties are then demonstrated in terms of the reflection and transmission behaviour of Gaussian beam solutions at an interface between two different metamaterials. We extend this treatment to N layered metamaterials using the Transfer Matrix Method. We demonstrate both positive and negative refraction and beam steering. Our proposed quantum graph modelling technique is very flexible and can be easily adjusted making it an ideal design tool for creating metamaterials with exotic band diagram properties or testing promising multi-layer set ups and wave steering effects.

**Metamaterials and quantum graphs.** Since Veselago published his pioneering paper in 1968—*The Electrodynamics of Substances with Simultaneously Negative Values of Permeability and Permittivity*[1]—it has been understood that the manipulation of electromagnetic material properties can give rise to exotic wave effects. In this paper, Veselago showed that when a material has both permittivity and permeability less than zero, its refractive index would be negative. By balancing the wave vector and the Poynting vector, it was shown that the electromagnetic waves within would have anti-parallel phase and group velocities. As a result, Snell's law and the Doppler and Vavilov–Cherenkov effects are reversed. These results were considered purely theoretical, since no such material was known to exist.

In 2000, Pendry demonstrated a remarkable application of negative refractivity. Showing, given an ideal lossless slab of material with a refractive index of $n = -1$ in a medium with equal and opposite refractive index, a lens can be theoretically constructed that could focus light perfectly[2]. Crucially such a material could overcome the diffraction limit of a traditional lens. Pendry presents a practical way one could engineer such a material. Proposing the periodic arrangement of unit cells made from C-shaped metal elements or "split-ring-resonators" with wires, giving rise to an effective negative refractive index within some frequency domain. This was experimentally demonstrated by Smith[3]. Since then, these man-made materials or "metamaterials" have been investigated extensively. There have been countless proposals for different resonator designs and arrangements, giving rise to a large number of different wave effects.

Metamaterials function due to the interplay between the wavelength and the scale of the unit cell. For wavelengths of the order or less than that of the unit cell, the waves undergo Bragg scattering interacting directly with each resonant element. However, in the long wavelength regime, the material appears continuous with properties owing to the underlying structure[4,5]. By varying these structural or resonant elements, one can achieve the desired wave effects. The required constituents of a metamaterial continue to be a matter of debate and various modelling techniques are used such as, transmission line models[6], boundary element models[7] and Finite Element Analysis[8] to gain insight into the design changes required to achieve the desired effects. These simulation techniques can be quite time consuming when setting up the models, let alone considering design modifications, so there is a need for simple and computationally cheap models for proof of principle studies. The proposed quantum graph approach can provide this simplified tool for designing metamaterial with a large range of different properties.

[1]School of Mathematical Sciences, University of Nottingham, Nottingham, NG7 2RD, UK. [2]Department of Mechanical Engineering and Mecha(tro)nic System Dynamics (LMSD), KU Leuven, 9000 Leuven, Belgium. ✉email: gregor.tanner@nottingham.ac.uk

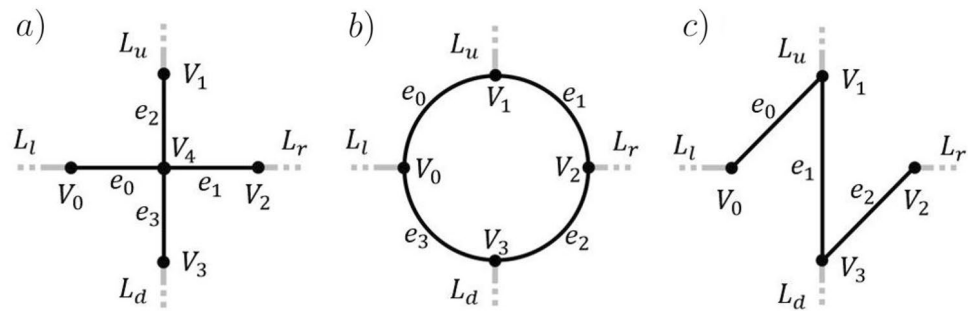

**Figure 1.** Three examples of open quantum graphs are shown. (**a**) Represents a cross resonator, (**b**) a ring resonator and (**c**) some zig zag structured resonator. The edges between vertices are depicted in black and open leads in grey.

Since their introduction in 1997[9], quantum graphs have become a most valuable model for studying quantum and wave effects in the context of quantum chaos and beyond. The theory of quantum graphs describe a network constructed from a set of vertices that act as point scatterers connected by a set of one-dimensional edges (bonds), wherein waves are free to travel. The wave transport through the network depends on the chosen properties of the point scatterers characterised by unitary scattering matrices. By changing the elements of the scattering matrix, one can customise the wave transport through the network. The versatility and ease of construction of quantum graphs, as well as the finite dimensionality and the exactness of 'semiclassical' expressions, make them ideal toy models to test ideas and research hypothesis'—see[10] and[11] for an overview. Besides quantum chaos, applications encompass modelling the vibrations of coupled plates[12], formulating quantum random walks[13,14] as well as quantum search algorithms[15]. One advantage of a quantum graph formulation is, that eigenvalue conditions can be written in terms of a secular equation involving the determinant of a matrix of finite dimension. Similarly, the scattering matrix, describing the properties of open quantum graphs, can be given in terms of a closed form expression involving finite dimensional matrices.

Naturally the concise language of quantum graph theory lends itself well to describing metamaterials. The lattice nodes act as scatterers representing sub-wavelength resonant elements and the edges model phase modulation between each resonant element. For this work, the metamaterial is modeled as an infinite square periodic quantum graph embedded in a 2D real space. A generalisation to 3D is straightforward by adding additional edges and vertices in the extra dimension. With this simple quantum graph construction, a multitude of non-trivial wave effects can be modelled by adjusting the parameters of the graph metric and the node scattering matrix. By manipulating the resulting band diagrams, we will demonstrate some of these effects, namely positive and negative refraction as well as beam steering.

The paper is structured as follows: in section "Waves in finite open quantum graphs—modelling resonant element" we will describe the set up of open quantum graphs which will serve as the resonant elements in the metamaterials. We will then construct period quantum graphs in section "A quantum graph model for a metamaterial" and derive the equations for constructing plane wave solutions and the dispersion curves in section "Wave propagation in metamaterials—plane wave solutions". In section "Wave propagation in metamaterials—Gaussian beams", we consider wave energy transport based on a Gaussian beam representation of the wave solutions and focus then on the available parameter space used for band engineering in section "Band engineering". In the last two sections, section "Wave refraction between metamaterials" and section "Waves in N layered metamaterials", we focus on reflection/transmission at "interfaces" between different metamaterials both in the two and N-layer case.

## Waves in finite open quantum graphs—modelling resonant element

Metamaterials are typically constructed from a periodic arrangement of sub-wavelength resonant elements. We consider here in particular two-dimensional metamaterials, a generalisation to three dimensions is straightforward. Using quantum graph theory, we can model a single resonant element as an open quantum graph, which is then periodically arranged to construct the metamaterial.

Consider a connected and open quantum graph $\Gamma(V, E, L)$ as a model for a resonator. The graph is constructed from a set of bidirectional edges $E = \{e_0, e_1, \ldots, e_{|E|-1}\}$ with a corresponding metric $\mathcal{L} = \{l_{e_0}, l_{e_1}, \ldots, l_{e_{|E|-1}}\}$, connected by a set of vertices $V = \{V_0, V_1, \ldots, V_{|V|-1}\}$ with corresponding boundary conditions. In addition, a set of four leads $L = \{L_l, L_r, L_d, L_u\}$ (semi-infinite edges) are imposed in the left($l$), right($r$), down($d$) and up($u$) directions serving as incoming and outgoing channels in the two dimensional lattice, see the examples in Fig. 1. Both the edges in $E$ and leads in $L$ are endowed with a one-dimensional wave equation,

$$\left(\frac{\partial^2}{\partial z^2} + k^2\right)\boldsymbol{\psi}(z) = 0. \tag{1}$$

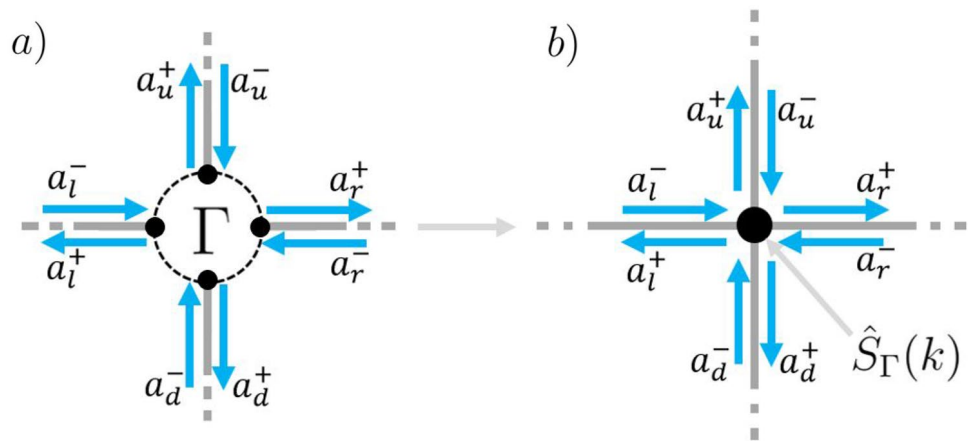

**Figure 2.** (**a**) Some arbitrary compact quantum graph $\Gamma(V, E, L)$. (**b**) The graph having been reduced to a single vertex. Noted are the lead wave amplitudes $\boldsymbol{a}_L^{\pm}$.

Here $\boldsymbol{\psi}(z) = (\boldsymbol{\psi}_L(z), \boldsymbol{\psi}_E(z))^T$ is the vector of all lead and edge solutions and $k$ is the wave number. The solution on a given edge $\psi_{e_j}(z)$ exists for $z \in [0, l_{e_j}]$, while the solution on a given lead $\psi_{L_j}(z)$ exists for $z \in [0, \infty)$. (For ease of notation, $z$ represents the spacial coordinates on any edge or lead.) The eigenfunction solution $\boldsymbol{\psi}(z)$ is expressed as a superposition of counter propagating plane waves,

$$\boldsymbol{\psi}(z) = \boldsymbol{a}^+ e^{ikz} + \boldsymbol{a}^- e^{-ikz}. \tag{2}$$

Here $\boldsymbol{a}^{\pm} = (\boldsymbol{a}_L^{\pm}, \boldsymbol{a}_E^{\pm})^T$ represents the vector of all wave amplitudes heading in $(-)$ or out $(+)$ of a vertex between the leads and edges. The vertices in $V$ are generally considered to be point scatterers, and the mapping from all incoming to all outgoing wave amplitudes at the vertices is described by a unitary matrix $\hat{S}$, that is,

$$\boldsymbol{a}^+ = \hat{S}\boldsymbol{a}^-. \tag{3}$$

We assume Neumann boundary conditions at individual vertices enforcing wave continuity and flux conservation at each vertex[10] although more general vertex scattering matrices can be considered[16]). Enforcing Neumann boundary conditions, we obtain for the $pq$th element of the scattering matrix associated with vertex $V_j$

$$\left\{\hat{S}_{V_j}\right\}_{pq} = \frac{2}{v_{V_j}} - \delta_{pq}. \tag{4}$$

Here, $\delta_{pq}$ is the Kronecker delta, and $v_{V_j}$ represents the valency (number of edges and leads attached to a given vertex) at vertex $V_j$.

To treat the connected graph as a model for a resonator, we wish to understand how waves incoming from any of the leads are redistributed and modulated into outgoing waves leaving through the leads. To do this a four dimensional graph scattering matrix $\hat{S}_\Gamma$ is constructed,

$$\boldsymbol{a}_L^+ = \hat{S}_\Gamma(k)\boldsymbol{a}_L^-, \tag{5}$$

reducing the entire connected graph in $\Gamma(V, E, L)$ to a single scatterer illustrated in Fig. 2.

To construct $\hat{S}_\Gamma$ from the underlying connected quantum graph, one decomposes the scattering matrix into discrete events. First, one accounts for the prompt reflections from the leads back to the leads in a matrix $\hat{S}_{LL}$. Second, the waves on the leads are coupled to the edges in $E$ with a matrix $\hat{S}_{EL}$. Third, the edge dynamics are described by an infinite series of scattered paths expressed as a Neumann series, $\sum_{n=0}^{\infty}(\hat{P}(k)\hat{S}_{EE})^n\hat{P}(k) := [\hat{\mathbb{1}} - \hat{P}(k)\hat{S}_{EE}]^{-1}\hat{P}(k)$. Here, $\hat{P}(k)$ maps the outgoing wave amplitudes on the edges $\boldsymbol{a}_E^+$ to the incoming wave amplitudes $\boldsymbol{a}_E^-$, taking account of the phase modulation owing to the metric $\mathcal{L}$, and $\hat{S}_{EE}$ represents the vertex scattering between edges. Fourth and finally, the waves on the edges are coupled back onto the leads using a matrix $\hat{S}_{LE}$. Explicitly, we can write

$$\hat{S}_\Gamma(k) = \hat{S}_{LL} + \hat{S}_{LE}\left[\hat{\mathbb{1}} - \hat{P}(k)\hat{S}_{EE}\right]^{-1}\hat{P}(k)\hat{S}_{EL}, \tag{6}$$

see Kottos and Smilansky[17] for details. The matrix terms in $\hat{S}_\Gamma$ can be deduced from the open graph scattering matrix $\hat{S}$ in (3) using the block-matrix representation,

$$\begin{pmatrix} \boldsymbol{a}_L^+ \\ \boldsymbol{a}_E^+ \end{pmatrix} = \begin{pmatrix} \hat{S}_{LL} & \hat{S}_{LE} \\ \hline \hat{S}_{EL} & \hat{S}_{EE} \end{pmatrix} \begin{pmatrix} \boldsymbol{a}_L^- \\ \boldsymbol{a}_E^- \end{pmatrix} \tag{7}$$

In the next step, we will use this graph based resonator to construct a periodic metamaterial.

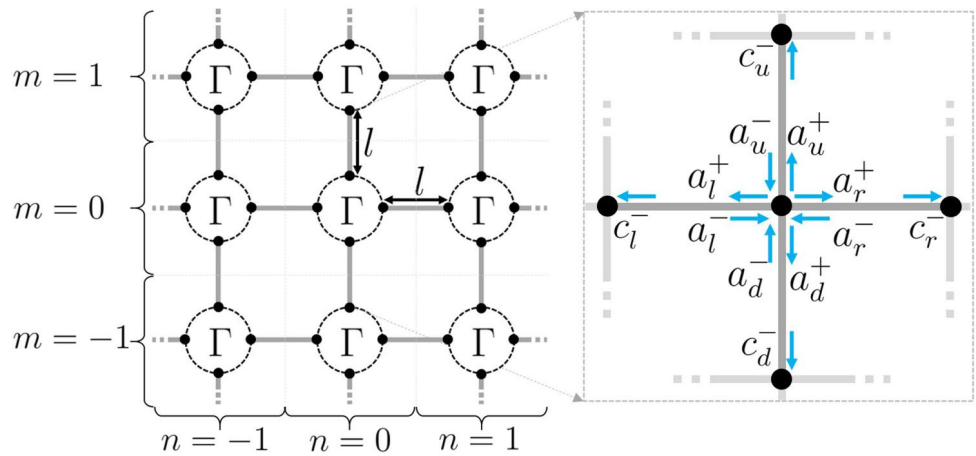

**Figure 3.** A 2D square periodic arrangement of quantum graph based resonant elements $\Gamma(V, E, L)$ connected by edges of length $l$ as a model for a 2D metamaterial. Location $n = m = 0$ shown in the subplot with labeled wave amplitudes, $\boldsymbol{a}^{\pm}$ and $\boldsymbol{c}^{-}$, traveling on the edges in the neighbourhood of the central vertex.

## A quantum graph model for a metamaterial

By periodically arranging these graph based resonant elements we will now model metamaterial behaviour. Consider placing each resonant element in a mesh, spaced by edges of length $l$. To keep track of the wave location in the graph, indices $nm$ are introduced, where $n$ representing steps in the horizontal direction and $m$ in the vertical, as illustrated in Fig. 3. We use $z$ again for the coordinate on the edges with $z = 0$ at a given resonant element and $z = l$ at the neighbouring element for all edge (where we leave out the indices $nm$ for convenience).

As before, all graph edges are endowed with the one dimensional wave equation (1). The vector of solutions is restricted to the edges $(l, r, d, u)$ in each unit cell of the material at location $nm$,

$$\boldsymbol{\psi}_{nm}(z) = \begin{pmatrix} \psi_{nm,l}(z) \\ \psi_{nm,r}(z) \\ \psi_{nm,d}(z) \\ \psi_{nm,u}(z) \end{pmatrix}. \tag{8}$$

To solve for the wave solution across the metamaterial, we begin by evaluating the scattering at the central vertex $n = m = 0$. Let $\boldsymbol{a}^{\pm} = (a_l^{\pm}, a_r^{\pm}, a_d^{\pm}, a_u^{\pm})^T$ be the vector of wave amplitudes heading in $(-)$ or out $(+)$ of the central vertex, see Fig. 3. We have

$$\boldsymbol{a}^{+} = \hat{\mathbb{S}}_{\Gamma}(k)\boldsymbol{a}^{-}. \tag{9}$$

The waves once scattered, undergo phase modulation as they move along each edge allowing one to express the outgoing waves at the central vertex $\boldsymbol{a}^{+}$ in terms of the incoming wave amplitudes $\boldsymbol{c}^{-} = (c_l^{-}, c_r^{-}, c_d^{-}, c_u^{-})^T$ at the neighboring vertices,

$$\boldsymbol{c}^{-} = e^{ikl}\boldsymbol{a}^{+}. \tag{10}$$

Due to the periodicity of the structure we can use Bloch's theorem to describe the wave solution in any unit cell $nm$ in terms of the solution at the central vertex $nm = 00$[18], that is,

$$\boldsymbol{\psi}_{nm}(z) = e^{i(\kappa_x n + \kappa_y m)l}\boldsymbol{\psi}_{00}(z). \tag{11}$$

Here, $\kappa_x$ and $\kappa_y$ are the Bloch wave numbers in the horizontal and vertical direction, respectively. The incoming wave amplitudes at a given vertex and its neighbours are thus related by

$$\boldsymbol{a}^{-} = \hat{B}(\kappa_x, \kappa_y)\boldsymbol{c}^{-} \tag{12}$$

where

$$\hat{B}(\kappa_x, \kappa_y) := \begin{pmatrix} 0 & e^{-i\kappa_x l} & 0 & 0 \\ e^{i\kappa_x l} & 0 & 0 & 0 \\ 0 & 0 & 0 & e^{-i\kappa_y l} \\ 0 & 0 & e^{i\kappa_y l} & 0 \end{pmatrix}. \tag{13}$$

By combining Eqs. (9), (10) and (12), we find the condition

$$\left[\hat{\mathbb{1}} - e^{ikl}\hat{B}(\kappa_x, \kappa_y)\hat{\mathbb{S}}_{\Gamma}(k)\right]\boldsymbol{a}^{-} = \boldsymbol{0}, \tag{14}$$

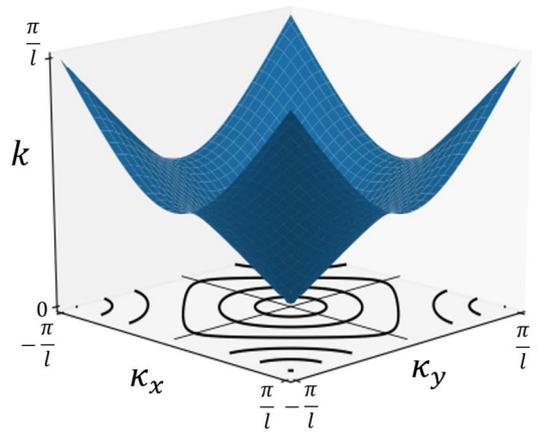

**Figure 4.** The band diagram for a lattice without resonant elements. Shown are iso-frequency contours for various values of $k$ in the $\kappa_x, \kappa_y$ plane.

which is satisfied if the wave vector $k$ on each edge and the quasi-wave vectors $\kappa_x$ and $\kappa_y$ fulfill the secular equation

$$\det\left[\hat{\mathbb{1}} - e^{ikl}\hat{B}(\kappa_x, \kappa_y)\hat{S}_\Gamma(k)\right] = 0. \tag{15}$$

In satisfying this condition, the incoming amplitudes at the central vertex $\boldsymbol{a}^-$ can be found by solving for the eigenvector in Eq. (14). We then obtain the wave solutions in each unit cell $nm$ throughout the metamaterial in the form

$$\boldsymbol{\psi}_{nm}(z; k, \kappa_x, \kappa_y) = e^{i(\kappa_x n + \kappa_y m)l}\left(\boldsymbol{a}^+(k, \kappa_x, \kappa_y)e^{ikz} + \boldsymbol{a}^-(k, \kappa_x, \kappa_y)e^{-ikz}\right), \tag{16}$$

where $\boldsymbol{a}^+$ is obtained via (9).

## Wave solutions for the infinite lattice

In the following, we will show how to construct plane wave and then beam-like solutions for the infinite lattice. The starting point for the construction are the solutions of Eq. (15) giving rise to dispersion curves of the form $k(\kappa_x, \kappa_y)$ or $\kappa_x(k, \kappa_y)$. We encounter real or imaginary solutions of $\kappa_x(k, \kappa_y)$ corresponding to propagating or evanescent waves, respectively, the latter occurring in band gaps. The full wave solution can then be constructed from the corresponding eigenvectors in (14) extended throughout the lattice using the Bloch condition (11), as formulated in (16).

### Wave propagation in metamaterials—plane wave solutions.

We start by constructing plane wave solutions and will discuss a particular simple case here, where the dispersion curves can be given analytically. That is, we consider a lattice without resonators where each edge is directly connected through the vertex. Here, the vertex acts as a point scatterer, where the scattering matrix $\hat{S}_\Gamma(k)$ in (6) is given by the Neumann scattering matrix (4) with $\nu_V = 4$, that is,

$$\hat{S}_{\Gamma_{point}} = \frac{1}{2}\begin{pmatrix} -1 & 1 & 1 & 1 \\ 1 & -1 & 1 & 1 \\ 1 & 1 & -1 & 1 \\ 1 & 1 & 1 & -1 \end{pmatrix}. \tag{17}$$

A generalisation to arbitrary $\hat{S}_\Gamma(k)$ is straightforward, examples will be shown in later sections. By solving Eq. (15), the dispersive properties of the lattice are given by the relation

$$2\cos(kl) = \cos(\kappa_x l) + \cos(\kappa_y l). \tag{18}$$

The resulting surface, or band, is shown in Fig. 4; it is periodic in $\boldsymbol{\kappa}$-space over the Brillouin Zone (*BZ*), where $BZ \in [-\frac{\pi}{l}, \frac{\pi}{l}]^2$. The shape of the band is a function of the square periodic graph topology and the symmetry of the scattering matrix. To generate a solution consider taking a single value of $k$, resulting in a discrete ring of $\boldsymbol{\kappa}$ solutions on an iso-frequency contour. By picking a point on the contour, one constructs the full wave solution from the corresponding eigenfunction in Eq. (16) as plotted in Fig. 5.

The phase of the resulting eigenfunction is given by the chosen Bloch wave vector $\boldsymbol{\kappa} = (\kappa_x, \kappa_y)^T$, while the energy flow is given by the Poynting vector $\boldsymbol{J} = (J_x, J_y)^T$ normal to the iso-frequency contour. To construct the components of $\boldsymbol{J}$ explicitly, one evaluates the flux on the graph edges. The 1D flux $J$ of the wave function $\boldsymbol{\psi}_{nm}$ on edge $p$ is

$$J(\psi_{nm,p}(z)) := \Re\left(\bar{\psi}_{nm,p}(z)\frac{1}{i}\frac{\partial \psi_{nm,p}(z)}{\partial z}\right) = k(|a_p^+|^2 - |a_p^-|^2) \tag{19}$$

with $\bar{\psi}$ denoting the complex conjugate of $\psi$. The horizontal and vertical components of the Poynting vector can then be evaluated in terms of the waves on edges right($r$)(or left($l$)) and up($u$)(or down($d$)), respectively, that is

$$\boldsymbol{J} = \begin{pmatrix} J_x \\ J_y \end{pmatrix} := k\begin{pmatrix} |a_r^+|^2 - |a_r^-|^2 \\ |a_u^+|^2 - |a_u^-|^2 \end{pmatrix} = k\begin{pmatrix} |a_l^-|^2 - |a_l^+|^2 \\ |a_d^-|^2 - |a_d^+|^2 \end{pmatrix}. \tag{20}$$

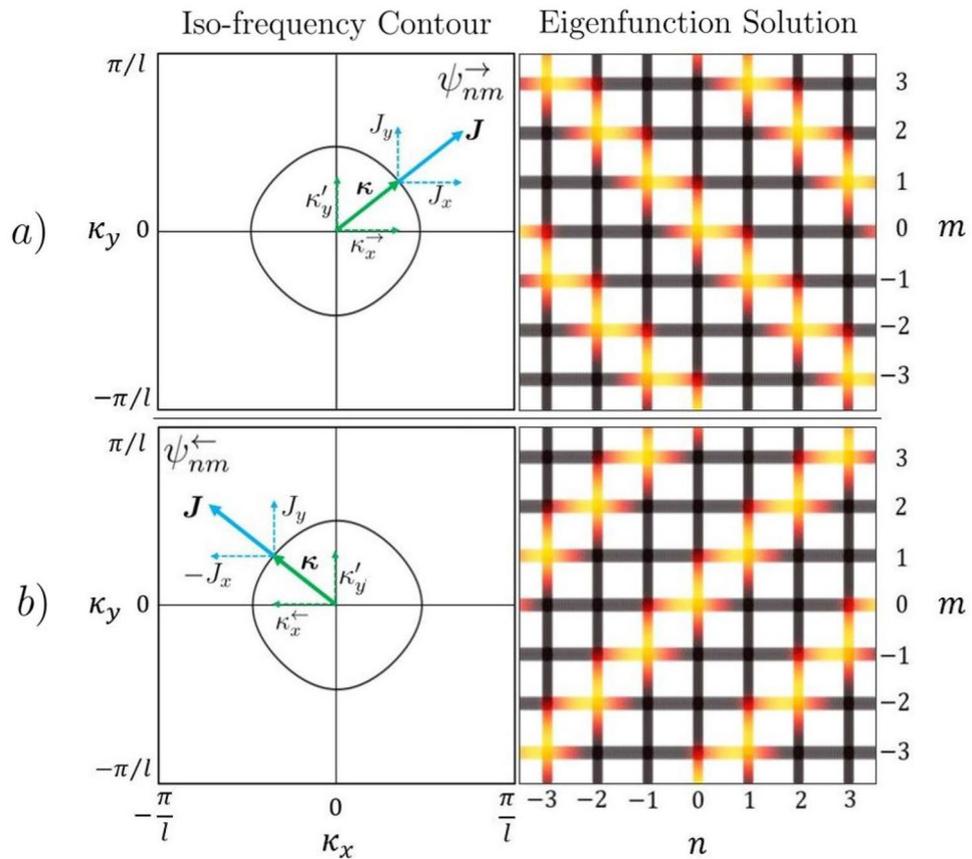

**Figure 5.** Iso-frequency contour of Fig. 4 with possible wave vectors $\boldsymbol{\kappa} = (\kappa_x^{\rightarrow}, \kappa_y')^T$ in (**a**) and $\boldsymbol{\kappa} = (\kappa_x^{\leftarrow}, \kappa_y')^T$ in (**b**), both shown in green, with corresponding Poynting vector $\boldsymbol{J} = (J_x, J_y)^T$ and $\boldsymbol{J} = (-J_x, J_y)^T$ normal to the contour shown in blue. (See next section for the choice of notation.) The resulting real components of the eigenfunction solutions $\boldsymbol{\psi}_{nm}^{\leftrightarrows}(z; k, \kappa_x, \kappa_y)$ for the values of the wave vectors used in (**a**) and (**b**) are also shown.

**A point on notation.** In the example shown in Fig. 5, for every value of $\kappa_y$, there are two corresponding values of $\kappa_x$. Naturally, the choice of $\kappa_x$ informs the direction of energy flow. To delineate between waves traveling in opposite horizontal directions, the following notation is used: eigenfunction solutions with Poynting vector heading to the right are given index $\rightarrow$ and the corresponding eigenvector in Eq. (14) is expressed as $\boldsymbol{a}^-$ as before where $\boldsymbol{a}^{\pm} := \boldsymbol{a}^{\pm}(k, \kappa_x^{\rightarrow}, \kappa_y)$; eigenfunction solutions with Poynting vector heading to the left are given index $\leftarrow$ with the corresponding eigenvector in (14) relabeled as $\boldsymbol{b}^-$ where $\boldsymbol{b}^{\pm} := \boldsymbol{a}^{\pm}(k, \kappa_x^{\leftarrow}, \kappa_y)$. Explicitly,

$$\boldsymbol{\psi}_{nm}(z; k, \kappa_x, \kappa_y) := \begin{cases} \boldsymbol{\psi}_{nm}^{\rightarrow}(z; k, \kappa_y) = e^{i(\kappa_x^{\rightarrow} n + \kappa_y m)l}\left(\boldsymbol{a}^+ e^{ikz} + \boldsymbol{a}^- e^{-ikz}\right), & J_x > 0 \\ \boldsymbol{\psi}_{nm}^{\leftarrow}(z; k, \kappa_y) = e^{i(\kappa_x^{\leftarrow} n + \kappa_y m)l}\left(\boldsymbol{b}^+ e^{ikz} + \boldsymbol{b}^- e^{-ikz}\right), & J_x < 0 \end{cases}. \tag{21}$$

In choosing the value of $\kappa_x$ ($\kappa_x^{\rightarrow}$ or $\kappa_x^{\leftarrow}$), one implicitly chooses the wave direction. For this reason the $\kappa_x$ dependence is dropped from the eigenfunction.

**Wave propagation in metamaterials—Gaussian beams.** Plane wave solutions travel in the direction of the wave vector $\boldsymbol{\kappa}$, which does not necessarily agree with the direction of energy flow given by the Poynting vector $\boldsymbol{J}$ in (20). This can be observed when considering Gaussian beam solutions constructed via Fourier series with the eigenfunctions, shown above, as a basis. The solution of the Gaussian beam with focal point at $n = n'$, expressed in components $\boldsymbol{\Phi}_{nm}(z) = (\Phi_{nm,l}(z), \Phi_{nm,r}(z), \Phi_{nm,d}(z), \Phi_{nm,u}(z))^T$, is given as,

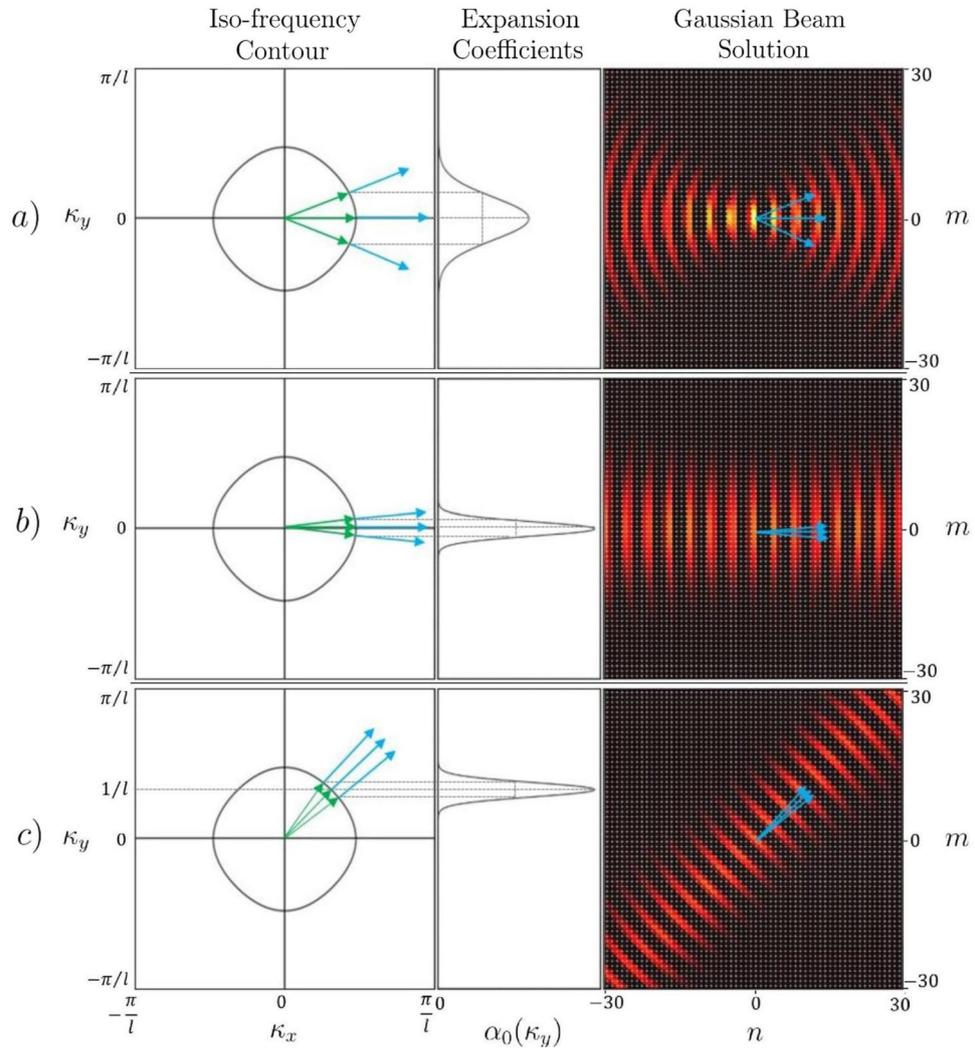

**Figure 6.** Iso-frequency contour for a metamaterial with resonant elements $\hat{S}_{\Gamma_{point}}$ as defined in (17) at $k = 1/l$ together with the expansion coefficients $\alpha_{n'=0}$ in Eq. (23) and the resulting real components of the Gaussian beam profile $\Phi_{nm}$ with focal point set to $n' = 0$. The beam profiles, Eq. (24), shown here are characterised by the parameters $\kappa'_y = 0$, 0 and $1/l$ and beam widths $\sigma = 2.2l$, $6.6l$ and $6.6l$ for (**a**), (**b**) and (**c**), respectively.

$$\Phi_{nm}(z; k) = \frac{1}{\sqrt{2\pi}} \int_\Omega \alpha_{n'}(\kappa_y; k) \vec{\psi}_{nm}(z, \kappa_y; k) d\kappa_y, \tag{22}$$

where the integral is performed over the domain $\Omega = \Omega(k)$ of the iso-frequency contour. Solutions outside $\Omega$ are evanescent and so contribute nothing in the far field. The expansion coefficients $\alpha_{n'}$ describing the beam profile at the focal point in terms of the eigenfunctions, given by the inverse transform,

$$\alpha_{n'}(\kappa_y; k) = \frac{1}{\sqrt{2\pi}} \sum_{m=-\infty}^{\infty} \left\{ \int_0^l \bar{\vec{\psi}}_{n'm,u}(z, \kappa_y; k) \Phi_{n'm,u}(z; \kappa_y) dz \right\}. \tag{23}$$

Here, $\bar{\vec{\psi}}_{n'm,u}$ is the complex conjugate of the right moving eigenfunction expressed for all $m$ at horizontal location $n = n'$ along the upward edges $u$ and $\Phi_{n'm,u}(z)$ is the beam profile for all $m$ with focal point $n = n'$ expressed on edge $u$. We chose the beam profile to be Gaussian, such that

$$\Phi_{n'm}(z) = \frac{1}{\sqrt{\sigma}\sqrt{\pi}} e^{-\left(\frac{z+ml}{\sqrt{2}\sigma}\right)^2} e^{i\kappa'_y l}, \tag{24}$$

where $\sigma$ is the width of the beam and the phase $e^{i\kappa'_y l}$ determines the tilt angle of the beam with respect to the horizontal axis of the lattice. Since the space on the graph is discretised, so too are the spacial integrals. Figure 6 illustrates how varying the parameters $\sigma$ and $\kappa'_y$ effect the shape and direction of the resulting Gaussian beam for a given iso-frequency contour $\Omega(k)$.

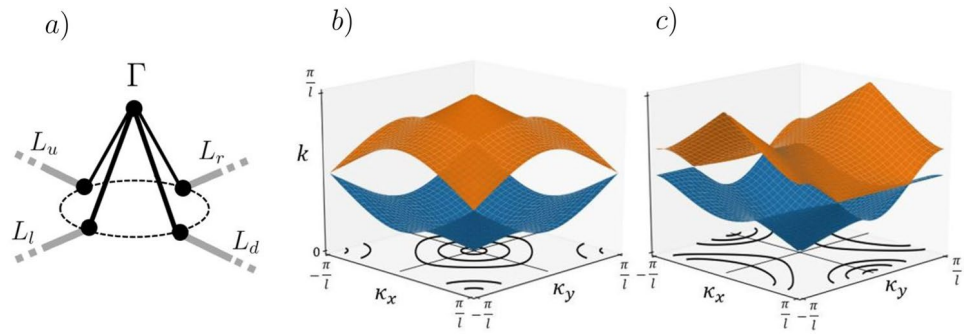

**Figure 7.** (**a**) Shows the cross resonator plugged into the metamaterials unit cell with scattering matrix defined as $\hat{S}_{\Gamma_{cross}}(k; l_x, l_y)$ in (25). Both (**b**) and (**c**) show the resulting band for different values of $l_x$ and $l_y$. Also plotted in the $\kappa_x, \kappa_y$ plane are the iso-frequency contours of the second band for various values of $k$. Plot (**b**) is for $l_x = 1$, $l_y = 1$. Plot (**c**) is for $l_x = 0, l_y = 1$.

## Band engineering

In a next step, we will use the quantum graph formulation for designing resonant elements leading to materials with "exotic" wave properties. As a simple example, we consider using the cross resonator illustrated in Fig. 1a as a resonant element for the metamaterial illustrated in Fig. 3. By setting the metric of the cross resonator to $\mathcal{L} = \{l_x/2, l_x/2, l_y/2, l_y/2\}$ as shown in Fig. 7a, one can modulate the phase in the horizontal and vertical direction across the resonant element thus breaking the scattering symmetry between the four leads. By enforcing Neumann boundary conditions at the vertices as defined by Eq. (4), the scattering matrix between the four leads $L$ has solution,

$$\hat{S}_{\Gamma_{cross}}(k; l_x, l_y) = \frac{1}{2} \begin{pmatrix} -e^{ikl_x} & e^{ikl_x} & e^{ik\left(\frac{l_x+l_y}{2}\right)} & e^{ik\left(\frac{l_x+l_y}{2}\right)} \\ e^{ikl_x} & -e^{ikl_x} & e^{ik\left(\frac{l_x+l_y}{2}\right)} & e^{ik\left(\frac{l_x+l_y}{2}\right)} \\ e^{ik\left(\frac{l_x+l_y}{2}\right)} & e^{ik\left(\frac{l_x+l_y}{2}\right)} & -e^{ikl_y} & e^{ikl_y} \\ e^{ik\left(\frac{l_x+l_y}{2}\right)} & e^{ik\left(\frac{l_x+l_y}{2}\right)} & e^{ikl_y} & -e^{ikl_y} \end{pmatrix}, \tag{25}$$

found by Eq. (6). By solving Eq. (15) the dispersion relation is now given as

$$\cos(\kappa_x l)e^{ik(l_x+l)}\left(e^{2ik(l_y+l)} - 1\right) + \cos(\kappa_y l)e^{ik(l_y+l)}\left(e^{2ik(l_x+l)} - 1\right) - e^{2ik(l_x+l_y+2l)} + 1 = 0. \tag{26}$$

By comparing Fig. 7b with 7c, we see that varying the phase modulation across each resonant element only in one dimension breaks the symmetry of the scattering and gives rise to a saddle shaped band, which can be exploited to yield negative refraction as shown in the next section.

## Wave refraction between metamaterials

To exemplify the engineered refractive properties, consider two metamaterials represented by semi infinite square periodic quantum graphs, one with resonant elements given by $\hat{S}_{\Gamma_1}$ and the other with resonant elements given by $\hat{S}_{\Gamma_2}$. The two materials are connected along the $y$ direction and material 1 exists for the set of unit cells $n_1 = (-\infty, \dots, -2, -1)$ and material 2 exists for the set of unit cells $n_2 = \{0, 1, \dots, \infty\}$, as illustrated in Fig. 8.

The full wave solution $\mathbf{\Psi}_{nm} = (\Psi_{nm,l}, \Psi_{nm,r}, \Psi_{nm,d}, \Psi_{nm,u})^T$ across the two materials can be constructed by a linear superposition of counter propagating eigenfunction solutions $\boldsymbol{\psi}_{j,nm}^{\rightleftharpoons}$ in material $j = 1$ and 2, as expressed in Eq. (21), that is,

$$\begin{aligned} \mathbf{\Psi}_{nm}(z, \kappa_y; k) = {} & H(n_1)\left[A_1(\kappa_y; k)\boldsymbol{\psi}_{1,nm}^{\rightarrow}(z, \kappa_y; k) + B_1(\kappa_y; k)\boldsymbol{\psi}_{1,nm}^{\leftarrow}(z, \kappa_y; k)\right] \\ & + H(n_2)\left[A_2(\kappa_y; k)\boldsymbol{\psi}_{2,nm}^{\rightarrow}(z, \kappa_y; k) + B_2(\kappa_y; k)\boldsymbol{\psi}_{2,nm}^{\leftarrow}(z, \kappa_y; k)\right], \end{aligned} \tag{27}$$

where $H(n_j)$ is the discretised Heaviside step function, that is,

$$H(n_j) = \begin{cases} 1, & \forall n \in n_j \\ 0, & \forall n \notin n_j \end{cases} \tag{28}$$

for $j = 1, 2$ and the coefficients $A_j$ and $B_j$ are associated with left and right moving waves, respectively. Thus, solutions with coefficients $A_1$ and $B_2$ represent waves incident on the interface, while solutions with coefficients $B_1$ and $A_2$ represent waves scattered from the interface. To determine the coefficients, we must satisfy the boundary conditions at the material interface. Naturally the wave solutions are given for a single value of $k$ between the two materials enforcing $k_1 = k_2 := k$. As the system stays periodic in the $y$ direction, the Bloch phase tangential to the interface also remain constant across the interface leading to the boundary condition $\kappa_{1,y} = \kappa_{2,y} := \kappa_y$. This condition is illustrated in Fig. 9 for a chosen value of $\kappa_y = \kappa'_y$ by a horizontal dashed grey line connecting the iso-frequency contours at $k = 1/l$ in material 1 and 2. The remaining unknowns are then the wave vectors normal to the interface $\kappa_{j,x}^{\rightleftarrows}(k, \kappa'_y)$ which can be obtained from the dispersion curve determined by solving Eq. (15) for each material, see the vertical dashed grey lines illustrated below.

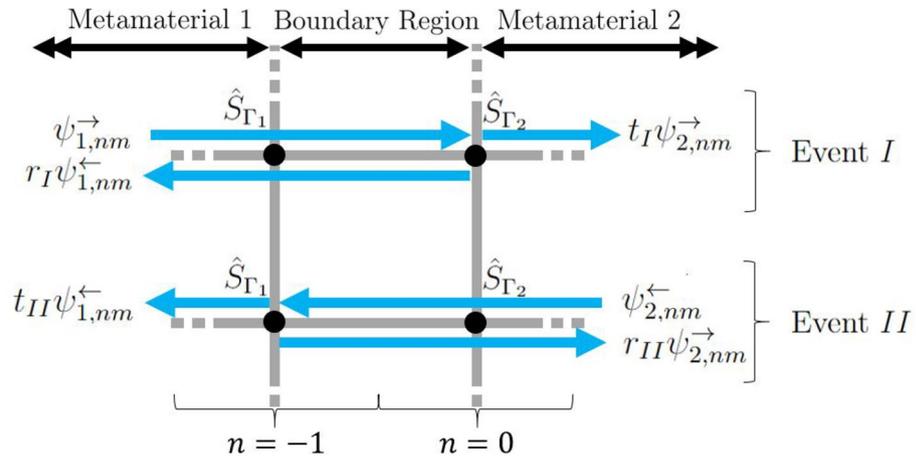

**Figure 8.** Boundary region between metamaterials 1 and 2, understood as all right($r$) edges for $n = -1$ and all left ($l$) edges for $n = 0$. Here, wave scattering from the boundary is divided into event $I$ and $II$, where $r_p$ and $t_p$ represent reflection and transmission amplitudes for event $p = I$ or $II$.

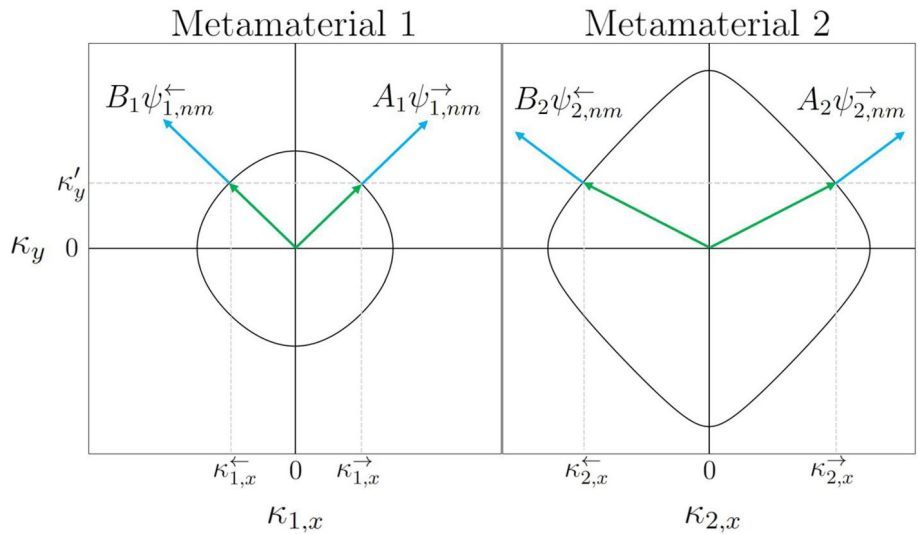

**Figure 9.** The iso-frequency contours for $k = 1/l$ for a scattering matrix with $\hat{S}_{\Gamma_1} := \hat{S}_{\Gamma_{point}}$ and $\hat{S}_{\Gamma_2} := \hat{S}_{\Gamma_{cross}}(k; 0, 1)$ representing the two metamaterials. The horizontal dashed grey line represents a single value of $\kappa_y = \kappa'_y$. Intersections with contour lines give the corresponding values of $\kappa_{j,x}^{\rightleftarrows}$ for $j = 1$ and 2.

To determine the scattering coefficients $A_j$ and $B_j$, we rescale the eigenfunction solutions of each material such that the magnitude of the horizontal flux of each component is equal, that is,

$$J(\psi_{1,nm,l}^{\rightarrow}) = J(\psi_{2,nm,l}^{\rightarrow}) = -J(\psi_{1,nm,l}^{\leftarrow}) = -J(\psi_{2,nm,l}^{\leftarrow}) \quad , \tag{29}$$

assuming $\kappa_{j,x}^{\rightleftarrows} \in \mathbb{R}$. Equally, the same condition can be enforced on the right edges, $r$. With the scaling choice (29), flux conservation across the interface

$$J(A_1\psi_{1,0m,l}^{\rightarrow} + B_1\psi_{1,0m,l}^{\leftarrow}) = J(A_2\psi_{2,0m,l}^{\rightarrow} + B_2\psi_{2,0m,l}^{\leftarrow}) \tag{30}$$

reduces to

$$|A_1|^2 + |B_2|^2 = |B_1|^2 + |A_2|^2 \tag{31}$$

and wave scattering at the interface can then be described in terms of a unitary scattering process. The corresponding interface scattering matrix $\hat{S}_{1,2}$ performing the mapping,

$$\begin{pmatrix} B_1 \\ A_2 \end{pmatrix} = \hat{S}_{1,2} \begin{pmatrix} A_1 \\ B_2 \end{pmatrix} \qquad (32)$$

can then be constructed by decomposing the interface scattering into two events. Event $I$ describes a wave incident from material 1 onto material 2 with amplitude $A_1 = 1$ and $B_2 = 0$, producing a reflected and transmitted wave with respective amplitudes $r_I$ and $t_I$. Event $II$ describes a wave incident from material 2 onto material 1 with amplitude $B_2 = 1$ and $A_1 = 0$, producing a reflected and transmitted wave with respective amplitudes $r_{II}$ and $t_{II}$, see Fig. 8. The interface scattering matrix takes on the form

$$\hat{S}_{1,2} = \begin{pmatrix} r_I & t_{II} \\ t_I & r_{II} \end{pmatrix}. \qquad (33)$$

To evaluate the matrix elements, consider first event $I$. For simplicity we choose to evaluate the waves at location $n = 0$, at coordinate $z = 0$ on edge $l$. Since the phase $\kappa_z$ is the same in both materials, it is sufficient to evaluate the solutions at $m = 0$. Here,

$$\Psi_{(\forall n \leq -1)0,l}(0)|_{n=0} = \left(a_{1,l}^+ + a_{1,l}^-\right) + r_I\left(b_{1,l}^+ + b_{1,l}^-\right) \qquad (34)$$
$$\Psi_{(\forall n \geq 0)0,l}(0)|_{n=0} = t_I\left(a_{2,l}^+ + a_{2,l}^-\right).$$

$\Psi_{(\forall n \leq -1)0,l}(0)|_{n=0}$ and $\Psi_{(\forall n \geq 0)0,l}(0)|_{n=0}$ both exist on the same edge in the boundary region, so to stay consistent, there must be an equivalence between the incoming($-$) and outgoing($+$) wave amplitudes of these solutions, that is,

$$t_I a_{2,l}^+ = a_{1,l}^+ + r_I b_{1,l}^+ \qquad (35)$$
$$t_I a_{2,l}^- = a_{1,l}^- + r_I b_{1,l}^-.$$

This can be solved to give

$$r_I = \frac{a_{2,l}^+ a_{1,l}^- - a_{2,l}^- a_{1,l}^+}{a_{2,l}^- b_{1,l}^+ - b_{1,l}^- a_{2,l}^+} \qquad (36)$$
$$t_1 = \frac{a_{1,l}^- b_{1,l}^+ - b_{1,l}^- a_{1,l}^+}{a_{2,l}^- b_{1,l}^+ - b_{1,l}^- a_{2,l}^+}.$$

Exactly the same procedure can be done for event $II$ where

$$\Psi_{(\forall n \leq -1)0,l}(0)|_{n=0} = t_{II}\left(b_{1,l}^+ + b_{1,l}^-\right) \qquad (37)$$
$$\Psi_{(\forall n \geq 0)0,l}(0)|_{n=0} = \left(b_{2,l}^+ + b_{2,l}^-\right) + r_{II}\left(a_{2,l}^+ + a_{2,l}^-\right),$$

which yields the equivalence condition

$$t_{II} b_{1,l}^+ = b_{2,l}^+ + r_{II} a_{2,l}^+ \qquad (38)$$
$$t_{II} b_{1,l}^- = b_{2,l}^- + r_{II} a_{2,l}^-$$

with solutions

$$r_{II} = \frac{b_{1,l}^+ b_{2,l}^- - b_{1,l}^- b_{2,l}^+}{a_{2,l}^+ b_{1,l}^- - a_{2,l}^- b_{1,l}^+} \qquad (39)$$
$$t_{II} = \frac{a_{2,l}^+ b_{2,l}^- - a_{2,l}^- b_{2,l}^+}{a_{2,l}^+ b_{1,l}^- - a_{2,l}^- b_{1,l}^+}.$$

The results of this unitary interface scattering is shown in Fig. 10 for a Gaussian beam incident from material 1 with amplitude $A_1 = 1$ and no incident beam from material 2, that is, $B_2 = 0$.

The Gaussian Beam is constructed based on the method described in section "Wave propagation in metamaterials—Gaussian beams", where the basis is the full wave field $\Psi_{nm}(x, \kappa_y; k)$ as defined in Eq. (27),

$$\Phi_{nm}(z) = \frac{1}{\sqrt{2\pi}} \int_\Omega \alpha_{-60}(\kappa_y) \Psi_{nm}(z, \kappa_y; k) d\kappa_y. \qquad (40)$$

The expansion coefficients $\alpha_{-60}$ are as in Eq. (23) for a beam focal point $n' = -60$ and tilt given by $\kappa_y' = 1/l$.

As was shown in section "Band engineering", breaking the symmetry of the scattering matrix gives rise to a saddle shaped band resulting in the property of negative refraction at appropriate $\kappa_y$ values, see Fig. 7c. When choosing such a material on the right hand side, one obtains the reflection/transmission behaviour shown in Fig. 11.

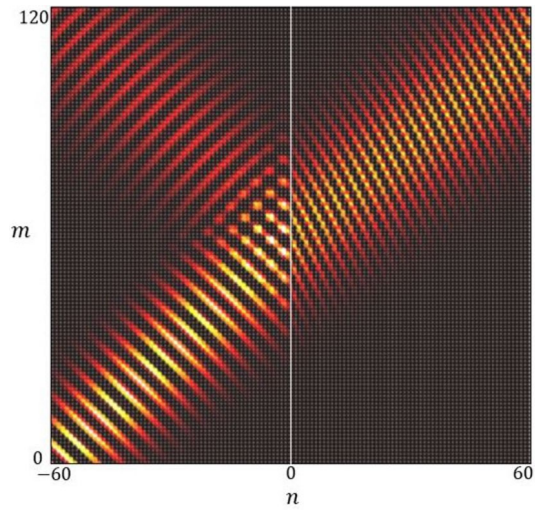

**Figure 10.** A refracted beam incident on a material interface with material properties defined by $\hat{S}_{\Gamma_1} := \hat{S}_{\Gamma_{point}}$ and $\hat{S}_{\Gamma_2} := \hat{S}_{\Gamma_{cross}}(k; 0, 1)$.

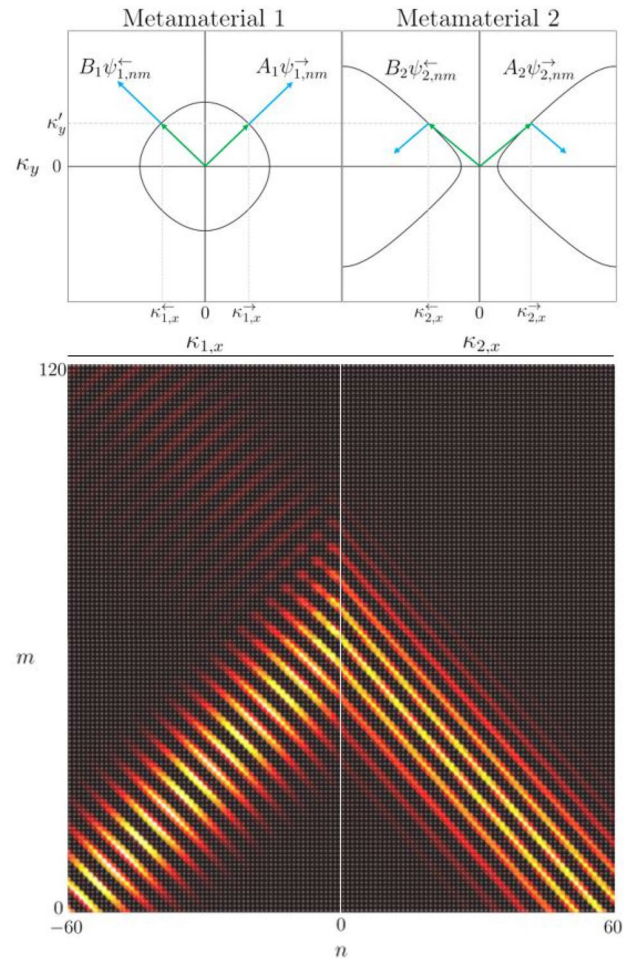

**Figure 11.** Top: Iso-frequency contours for $k = 1/l$ for a scattering matrix with $\hat{S}_{\Gamma_1} := \hat{S}_{\Gamma_{point}}$ and $\hat{S}_{\Gamma_2} := \hat{S}_{\Gamma_{cross}}(k; 0, 4.45)$. Bottom: The resulting real component of the Gaussian beam $\mathbf{\Phi}_{nm}$ constructed from the full wave field $\mathbf{\Psi}_{nm}$ across the two materials with incident wave amplitudes $A_1 = 1$ and $B_2 = 0$. The focal point of the Gaussian Beam is set to $n' = -60$ with tilt given by $\kappa'_y = 1/l$.

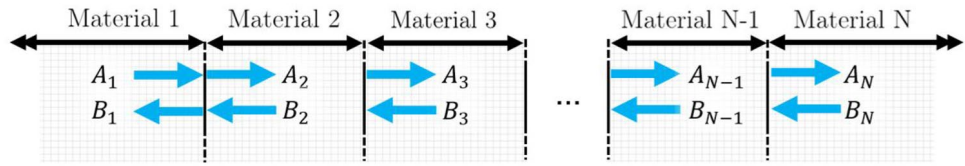

**Figure 12.** A system of $N$ layered metamaterials with wave amplitudes, $A_j$ and $B_j$, noted at each interface.

## Waves in *N* layered metamaterials

Consider now a system of $N$ layered metamaterials as shown in Fig. 12, each with their own properties as defined previously.

Each material spans a domain defined by the set of unit cells $n_j$. Here the wave function across all $N$ materials is expressed as a linear superposition of counter propagating eigenfunction solutions of each material, that is,

$$\Psi_{nm}(x, \kappa_y; k) = \sum_{j=1}^{N} H(n_j) \left[ A_j(\kappa_y; k) \overrightarrow{\psi}_{j,nm}(x, \kappa_y; k) + B_j(\kappa_y; k) \overleftarrow{\psi}_{j,nm}(x, \kappa_y; k) \right]. \tag{41}$$

To evaluate the full wave function across all $N$ materials, one must determine the coefficients $A_j$ and $B_j$ that satisfy the boundary conditions at all material interfaces. This is done by the Transfer Matrix Method[19]. In the previous section, the wave amplitudes between two materials were determined by constructing a scattering matrix mapping incoming to outgoing wave amplitudes between the metamaterials 1 and 2 as defined in Eq. (32). This matrix can be rearrange to give a transfer matrix that maps wave amplitudes from material 1 to material 2, $\hat{\tilde{S}}_{1,2}$.

$$\binom{A_2}{B_2} = \hat{\tilde{S}}_{1,2} \binom{A_1}{B_1} = \begin{pmatrix} t_I - \frac{r_I r_{II}}{t_{II}} & \frac{r_{II}}{t_{II}} \\ -\frac{r_I}{t_{II}} & \frac{1}{t_{II}} \end{pmatrix} \binom{A_1}{B_1}. \tag{42}$$

This procedure can be generalised to any arbitrary materials $j$ and $j + 1$ giving

$$\binom{A_{j+1}}{B_{j+1}} = \hat{\tilde{S}}_{j,j+1} \binom{A_j}{B_j}. \tag{43}$$

Waves propagating across a given material $j$ accumulate a Bloch phase which can be expressed in terms of the matrix

$$\hat{P}_j(\kappa_y; k) = \begin{pmatrix} e^{i\kappa_{j,x}^{\rightarrow}(\kappa_y; k) W_j} & 0 \\ 0 & e^{i\kappa_{j,x}^{\leftarrow}(\kappa_y; k) W_j} \end{pmatrix}, \tag{44}$$

where $W_j = (\max(n_j) - \min(n_j))l$ is the width of material $j$. Having now formulated both scattering and propagation, one can express the wave amplitudes in material $N$ in terms of the wave amplitudes in material 1.

$$\binom{A_N}{B_N} = \hat{\tilde{S}}_{N-1,N} \left( \hat{P}_{N-1} \hat{\tilde{S}}_{N-2,N-1} \right) \dots \left( \hat{P}_3 \hat{\tilde{S}}_{2,3} \right) \left( \hat{P}_2 \hat{\tilde{S}}_{1,2} \right) \binom{A_1}{B_1} := \hat{\tilde{S}}_{1,N} \binom{A_1}{B_1}. \tag{45}$$

We can now rearrange this transfer operator $\hat{\tilde{S}}_{1,N}$ such that the entire system of layered materials acts as a single point scatterer by introducing the scattering matrix $\hat{S}_{1,N}$ defined as

$$\binom{B_1}{A_N} = \hat{S}_{1,N} \binom{A_1}{B_N}. \tag{46}$$

Now by setting the incident wave amplitudes $A_1$ and $B_N$, we determine the amplitudes of the scattered field by direct substitution into Eq. (46). Knowing the wave amplitudes in material 1, $A_1$ and $B_1$, it is trivial to determine all other amplitudes using $\hat{P}_{j-1} \hat{S}_{j,j+1}$. The results of this procedure are plotted in Fig. 13 for a three layered material.

## Conclusion

We propose here a graph based technique for designing metamaterials allowing for a fast and flexible way to test resonant element proposals embedded periodically within a 2D square lattice. By modelling resonant elements in terms of open, graph-based scattering systems, we retain a connection with the underlying geometrical structure of the element which will inspire the engineering of physical metamaterials. This will make it possible to search for exotic band-diagrams, dispersion curves and wave effects linking these to an underlying array of resonant scattering systems. In this paper, we introduce the principles of graph based metamaterial construction including the set-up and we show how to obtain plane wave and beam-like solutions in the infinite medium. We then demonstrate the handling of boundary conditions at interfaces as well as the generalisation to $N$ layered media. In each case, the computation can be reduced to low-dimensional matrix problems with the help of Bloch's theorem. Further results modelling various special wave effects will be presented in forthcoming publications.

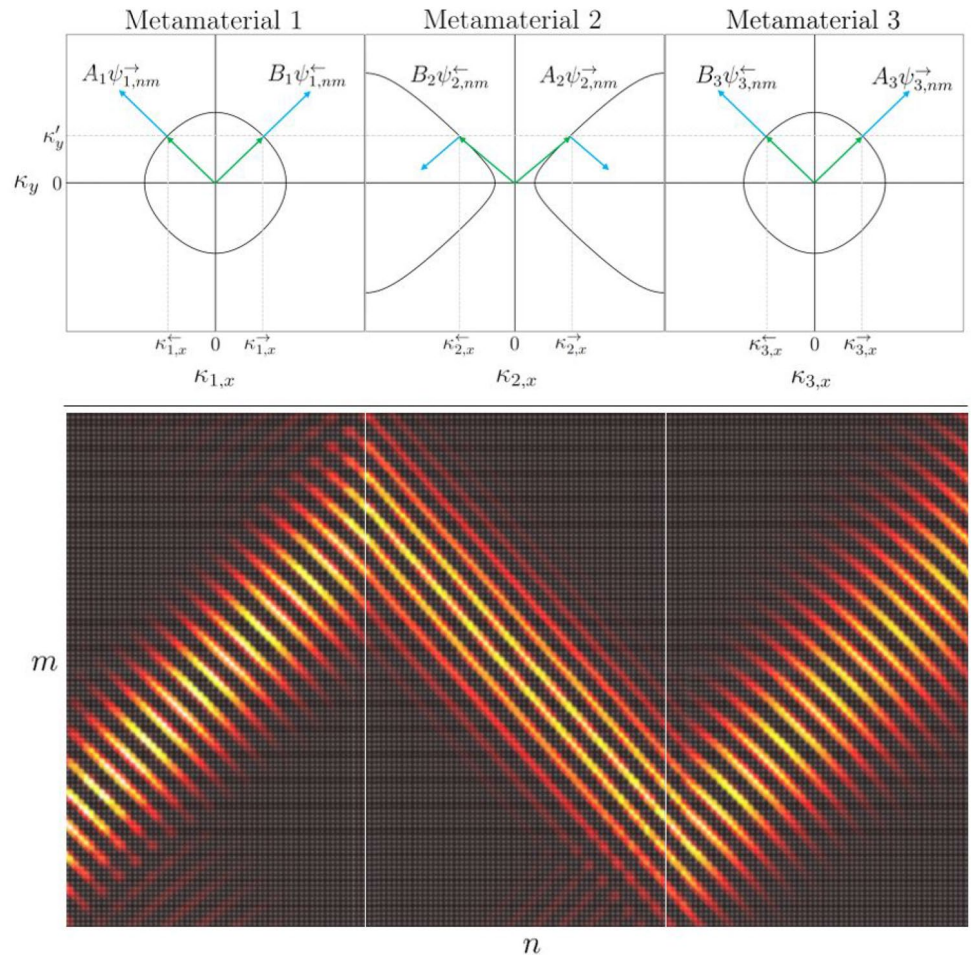

**Figure 13.** Top: The iso-frequency contours of three metamaterials with properties defined by scattering matrices $\hat{S}_{\Gamma_{point}}$, $\hat{S}_{\Gamma_{cross}}$ ($k$; 0, 4.45) and $\hat{S}_{\Gamma_{point}}$ for $k = 1/l$. Bottom: The real component of a Gaussian beam $\mathbf{\Phi}_{nm}$ incident from material 1 constructed from the full wave field $\Psi_{nm}$ for an incident wave from metamaterial 1, $A_1 = 1$, and no incident wave from material 3, $B_3 = 0$.

## Data availability

The datasets used and analysed during the current study are available from the corresponding author on reasonable request.

## Code availability

The code used to generate the datasets shown in this study are available from the corresponding author on reasonable request.

## Acknowledgements

The authors would like to thank Martin Richter, Gabriele Gradoni and Christian Blackman for interesting discussions and helpful comments. Their encouragement and support has been of great value to this work. This work has been supported by the EU H2020 RISE-6G project under grant number 101017011.

## Author contributions
T.L. wrote the code and carried out the numerical calculations. The manuscript has been written by T.L. and G.T. All authors took part in the scientific discussions and in proof reading the manuscript.

## Competing interests
The authors declare no competing interests.

## Additional information
**Correspondence** and requests for materials should be addressed to G.T.

**Reprints and permissions information** is available at www.nature.com/reprints.

**Publisher's note** Springer Nature remains neutral with regard to jurisdictional claims in published maps and institutional affiliations.